\documentclass[aps,twocolumn,prb,groupedaddress]{revtex4}
\usepackage{color,graphicx}
\usepackage{amsmath}
\usepackage{natbib}
\usepackage{longtable}

\begin{document}

\title{Lattice and spin excitations in multiferroic h-YMnO$_3$}

\author{C. Toulouse, J. Liu, Y. Gallais, M-A. Measson, A. Sacuto and M. Cazayous}
\affiliation{Laboratoire Materiaux et Phenomenes Quantiques UMR 7162 CNRS, Universite Paris Diderot-Paris 7, 75205 Paris cedex 13, France}
%\author{P. Rovillain}
%\affiliation{School of Physics, University of New South Wales, Sydney, New South Wales 2052, Australia}
%\affiliation{The Bragg Institute, ANSTO, Kirrawee DC NSW 2234, Australia}

\author{L. Chaix}

\affiliation{Institut Laue Langevin, 6 rue Jules Horowitz, BP 156, F-38042 Grenoble Cedex 9, France}
\altaffiliation{Institut N\'{e}el, CNRS et Universit\'{e} Joseph Fourier, BP166, F-38042 Grenoble Cedex 9, France}

\author{ V. Simonet and S. de Brion}
\affiliation{Institut N\'{e}el, CNRS et Universit\'{e} Joseph Fourier, BP166, F-38042 Grenoble Cedex 9, France}

\author{L. Pinsard-Godart}
\affiliation{Laboratoire de Chimie des Solides, URA 446 CNRS, Universite Paris-Sud, Orsay, France}

\author{F. Willaert, J. B. Brubach and P. Roy}
\affiliation{Synchrotron SOLEIL, L'Orme des Merisiers Saint-Aubin, BP 48, F-91192 Gif-sur-Yvette Cedex, France}

\author{S. Petit}
\affiliation{Laboratoire L\'{e}on Brillouin, CEA-CNRS, UMR 12, CE-Saclay, F-91191 Gif-sur-Yvette, France}

\date{\today}

\begin{abstract}
We used Raman and terahertz spectroscopies to investigate lattice and magnetic excitations and their cross-coupling in the hexagonal YMnO$_3$ multiferroic.
Two phonon modes are strongly affected by the magnetic order. Magnon excitations have been identified thanks to comparison with neutron measurements and spin wave calculations  but no electromagnon has been observed. In addition, we evidenced  two additional Raman active peaks. We have compared this observation with the anti-crossing between magnon and acoustic phonon branches measured by neutron. These optical measurements underly the unusual strong spin-phonon coupling.
\end{abstract}

%\pacs{77.80.Bh, 75.50.Ee, 75.25.+z, 78.30.Hv}
\maketitle

\section{Introduction}

Multiferroics combine two or more of the properties of (anti)ferromagnetism, ferroelectricity and ferroelasticity that can be coupled. The coupling between the magnetic and ferroelectric order parameters leads to magnetoelectric effect and the possibility to control magnetization by an electric field and vice versa.\cite{Eerenstein2006,Chu}
These materials are currently the subject of intensive investigations both because of the interesting physics involved and their potential applications in data storage, spintronics and sensors.\cite{Yang2007, Tokura2006, Kruglyak}
Among the multiferroic materials, RMnO$_3$ manganites have attracted a great deal of attention due to the significant coupling between the magnetic and electric order parameters. In orthorhombic manganites the magnetic frustrations leads to spin lattice coupling induced by the inverse Dzyaloshinski-Moriya interaction\cite{Cheong, Sergienko} whereas in hexagonal manganites the ferroelectric and magnetic orders are not induced by the same interaction\cite{Aken2004, Lee2008, Pimenov2006}.
YMnO$_3$ is one of the most studied hexagonal manganites due to evidences of the strong interplay between the magnetic and ferroelectric order.
The dielectric constant reveals clear anomalies at the N$\acute{e}$el temperature (\textsl{T$_N$}) and the electric moment is enhanced below T$_N$.\cite{Huang, Iwata, Yen, Lee}
 At the magnetic transition, large atomic displacements have been measured by diffraction techniques which reflects a strong magneto-elastic coupling.\cite{Lee2} In addition, the coupling between magnetic and ferroelectric domains has been imaged by optical second harmonic technique.\cite{Fiebig} Various experiments as thermal conductivity or Raman scattering have shown an unusually strong spin-lattice coupling in this compound.\cite{Cruz, Sharma, Litvinschuk} More recently, polarized inelastic neutron scattering measurements have  evidenced a hybrid boson mode in YMnO$_3$.\cite{Petit, Pailhes} All these measurements illustrate the crucial role played by the strong coupling between lattice, electric and magnetic degrees of freedom.

In this article, we used Raman and teraHerz (THz) spectrocopies to shed light on the spin-lattice coupling in the h-YMnO$_3$ multiferroic.
The phonon mode of the apical oxygen atom is affected by the magnetic transition showing that super-superexchange Mn-Mn interaction is involved in the stabilization of the three dimensional magnetic structure. The phonon mode associated to the apical yttrium ion is also sensitive to the magnetic order underlying the role of the magneto-elastic coupling between R and Mn ions in the h-RMnO$_3$ compounds.
We provide evidence of a strong coupling between magnon and phonon modes resulting in an anti-crossing between the dispersion of the acoustic phonon modes polarized along the ferroelectric axis and the magnon branches.
%We provide evidences of a strong coupling between magnon and phonon modes with the opening of a gap in the magnetic phase in the dispersion of the acoustic phonon modes polarized along the ferroelectric axis. %However, no hybrid excitation called electromagnon has been detected.

\section{Experimental Details}

\begin{figure}
 \includegraphics*[width=9cm]{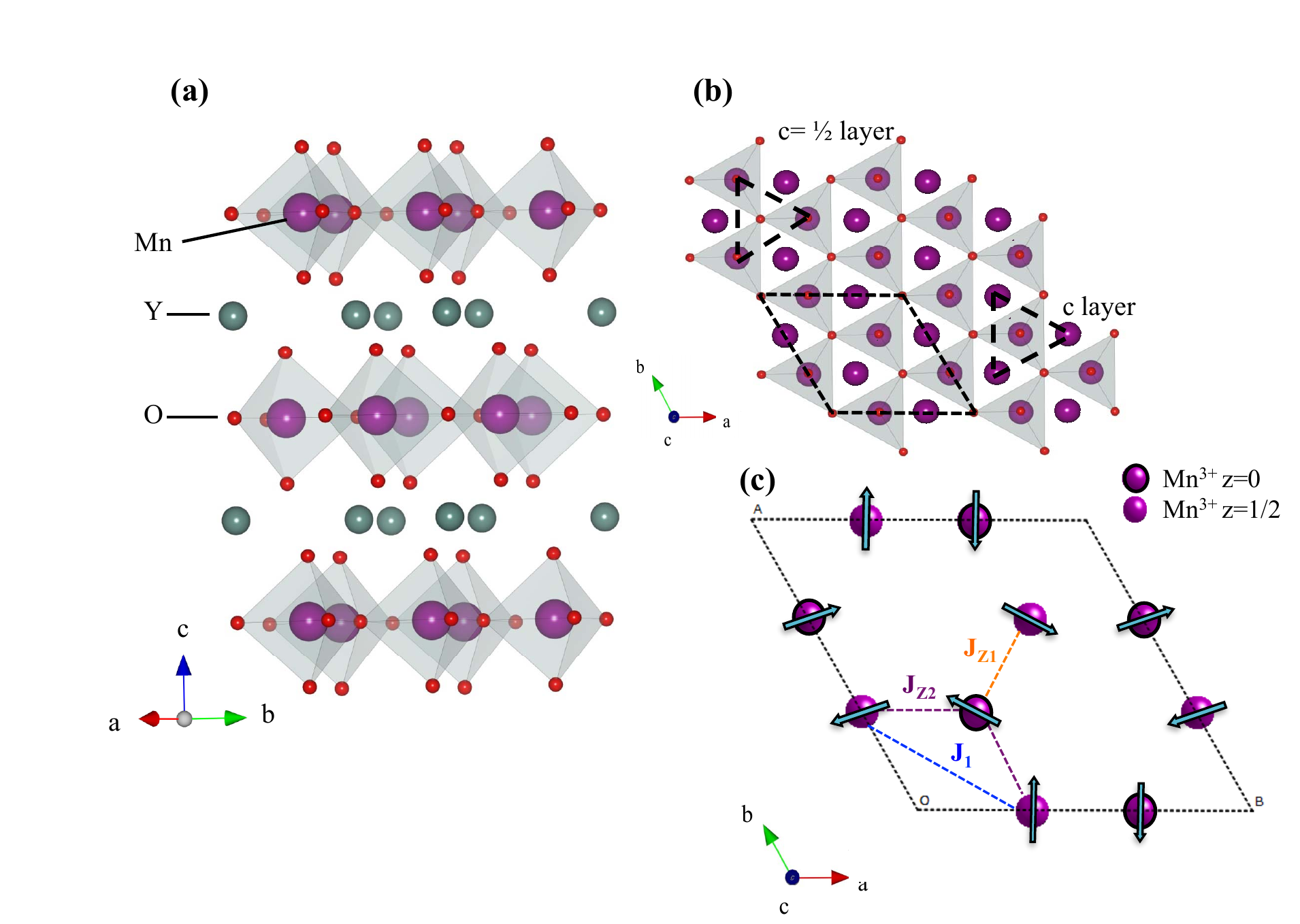}
 \caption{\label{Fig1} 	
 Crystallographic and magnetic structure of hexagonal YMnO$_3$ in its ferroelectric P6$_3$cm phase. The hexagonal unit cell is shown with dash lines. Panel (a) and (b) perspective and top views. Panel (c) hexagonal unit cell projection in the (a,b) plane with only the Mn$^{3+}$ ions. The corresponding $\Gamma_1$ magnetic order is also shown.}
\end{figure}

YMnO$_3$ crystallizes in the hexagonal symmetry (P6$_3$\textit{cm}) with lattice parameters equal to $a=6.15~\mathring{A}$ and $c=11.40~\mathring{A}$, even if the exact crystalline and magnetic structures are still under debate\cite{Singh2013}.
This compound is formed by stacked Mn-O and Y-O layers as shown in Fig. \ref{Fig1}. The Mn ions are surrounded
by three in-plane and two apical oxygen ions  and form two stacked triangular lattices.
The ferroelectric order appears below \textsl{T$_C$}$\approx$900~K resulting from the tilting of the MnO$_5$ bipyramides with the buckling of the Y-O planes. The polarization is along the c axis.
As a result, the oxygen ions move closer to the yttrium atoms giving rise to staggered ferroelectric
moments. YMnO$_3$ becomes antiferromagnetic below the N\'eel temperature \textsl{T$_N$}=72~K. The Mn magnetic moments order in 120$^o$ arrangements within the $\Gamma_1$ irreductible representation as determined by neutron diffraction measurements \cite{Fabreges2009,Fabreges2010}.
Magnetic frustration arises from the competition of the first neighbor antiferromagnetic interactions $J_1$ between the Mn$^{3+}$ spins in the triangular lattices (Fig. \ref{Fig1}).
%%The intraplane magnetic interactions are related to the superexchange interaction. However, the x coordinate of the Mn atoms is not exactly 1/3 in the (a,b) plane leading to two different magnetic interactions associated with two different Mn-Mn distances. The c and c=+1/2 layers are antiferromagnetically coupled by the supersuperexchange interaction through the apical oxygen atoms.

YMnO$_3$ single crystals have been grown using the standard floating zone technique. Several mm size plaquettes have been used with the c axis either in or perpendicular to the plaquette surface. The crystals have been polished to obtain high surface quality for Raman measurements while thickness of 600 $\mu$m have been used for transmission THz measurements.

Raman spectra were recorded in a backscattering geometry with a triple spectrometer Jobin Yvon T64000 coupled to a liquid-nitrogen-cooled CCD detector using the 514 excitation line from a Ar$^+$-Kr$^+$ mixed gas laser. The resolution of the excitation mode frequencies is less than 0.5 cm$^{-1}$. Temperature measurements have been performed using an ARS closed-cycle He cryostat and the magnetic field measurements using an Oxford Spectromag split-coil magnet.

THz absorbance spectra were obtained by measuring the transmission at the AILES (Advanced Infrared Line Exploited for Spectroscopy) beamline of Synchrotron SOLEIL.\cite{Soleil}
A Bruker IFS125 interferometer equipped with a pulse tube cryostat was used, combined with a Helium pumped bolometer.
The 10-60 cm$^{-1}$ energy range was explored at a resolution of 0.5 cm$^{-1}$ using a 6 $\mu$m thick silicon-mylar multilayered beamsplitter. The absolute absorbances
were determined by measuring the transmission through a 2 mm diaphragm as a reference and the sample transmission through that same diaphragm.

\section{Results and discussion}

\subsection{Lattice excitations}

The group theoretical analysis for the $\Gamma$-point phonon modes of hexagonal (P6$_3$\textit{cm}) YMnO$_3$ gives 60 phonon modes at the $\Gamma$-point: $10\textsl{A}_1+5\textsl{A}_2+10\textsl{B}_1+5\textsl{B}_2+15\textsl{E}_1+15\textsl{E}_2$ and 38 of these modes are Raman-active\cite{Iliev1997}: $\Gamma_{Raman} = 9\textsl{A}_1+14E_1+15\textsl{E}_2$.
Our measurements have been performed in backscattering configuration with the incident wave vector of the light anti-parallel to the scattered wave vector.
Pure \textsl{E}$_2$ modes are obtained using \textsl{z(xy)\={z}} geometry\cite{Porto1966} (corresponding to the backscattering configuration along the z-axis with polarization of the incident and scattered light along the x-axis and the y-axis respectively).
The \textsl{A}$_1$ modes are deduced from parallel polarizations with the \textsl{z(xx)\={z}} configuration giving the \textsl{A}$_1$(TO) + \textsl{E}$_2$ modes.

\begin{figure}
 \includegraphics*[width=7cm]{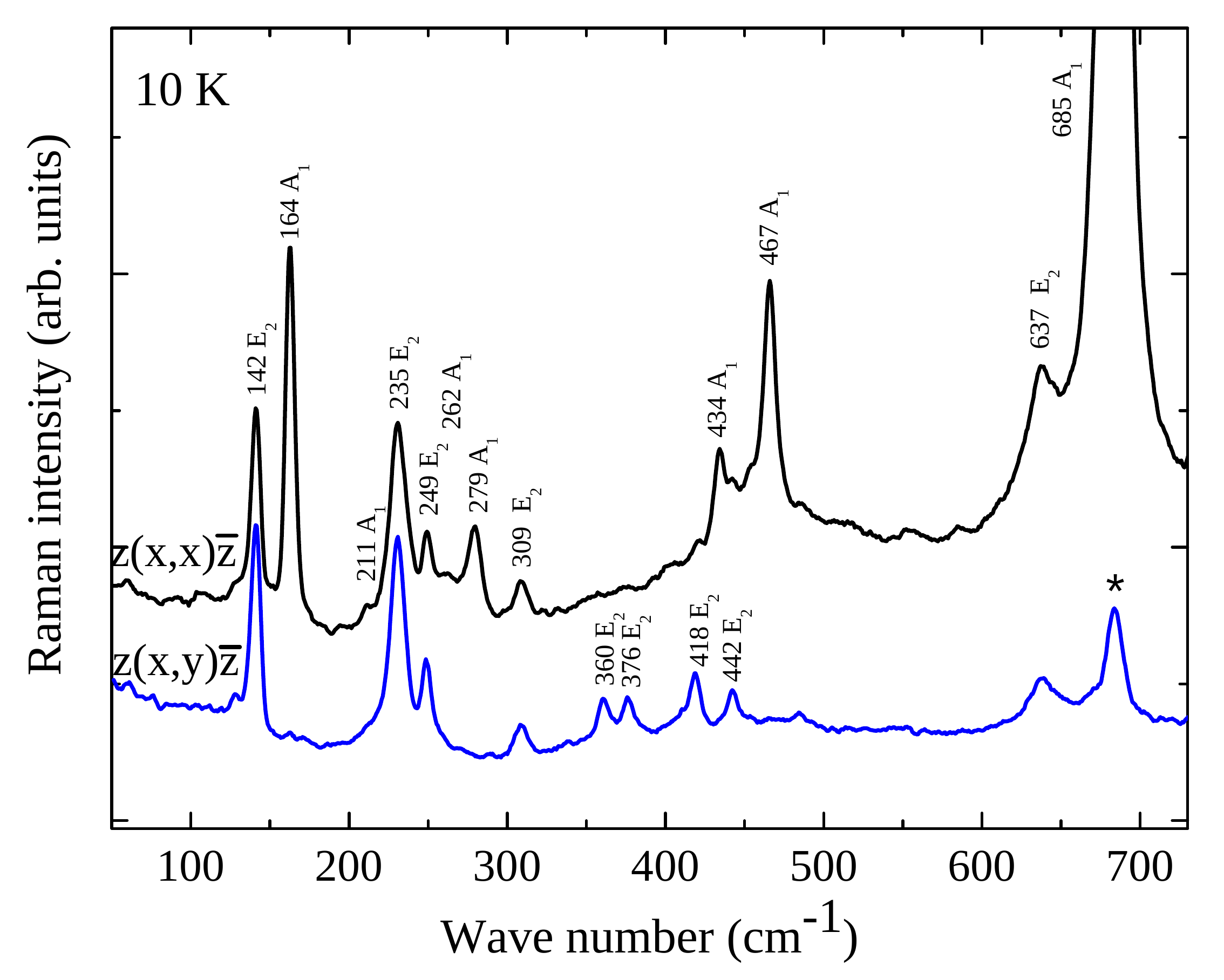}
 \caption{\label{Fig2} 	
Raman spectra of YMnO$_3$ single crystal measured at 10 K using a) \textsl{z(xx)\={z}} and b) \textsl{z(xy)\={z}} scattering configurations. Star indicates phonon mode due to polarization leakage.}
\end{figure}

Figure \ref{Fig2} shows the Raman spectra measured on h-YMnO$_3$ single crystals with \textsl{z(xx)\={z}} and \textsl{z(xy)\={z}} scattering configurations.
We have identified 7 \textsl{A}$_1$ modes and 9 \textsl{E}$_2$ modes. The frequencies of the phonon modes at 10~K are reported in Table I and compared to the previous experimental results on single crystals in addition to the associated atomic displacements.\cite {Iliev1997}

\begin{table*}[ht]
  \begin{center}
  \parbox{8cm}
  \caption{

    \textsl{A}$_1$ and \textsl{E}$_2$ mode frequencies (cm$^{-1}$) measured in \textsl{h}-YMnO$_3$ and description of the atomic displacements.}\\
	\vspace{3mm}
	 \begin{tabular}{cccccc}
	\hline\noalign{\smallskip}
	    Mode & This work & Iliev et al.\cite{Iliev1997} & Kim et al.\cite{Kim} &  Vermette et al.\cite{Vermette2010} & Direction of the largest \\
	       & 10 K & 300 K & 300 K & 10 K & displacement\cite{Iliev1997}\\
	\hline\noalign{\smallskip}
	  \textsl{A}$_1$ & 164 & 148 &     & 161 &+Z(Y$_1$), -Z(Y$_2$)  \\
	        & 211 & 190 & 205 & 244 &Rot x,y (MnO$_5$)    \\
	        & 262 & 257 &     & 264 &+Z(Y$_1$, Y$_2$), -Z(Mn)  \\
	        & 279 & 297 &     & 307 &X(Mn), Z(O$_3$)  \\
%	        &     &     &     &     &+Z(O$_3$),- Z(O$_4$), +X,Y(O$_2$), -X,Y(O$_1$)  \\
	        & 434 & 433 & 438 & 434 &+Z(O$_4$,O$_3$), -Z(Mn) \\
	        & 467 & 459 &     & 467 &+X,Y(O$_1$, O$_2$), -X,Y(Mn)  \\
%	        &     &     &     &     &+Z(O$_1$,O$_2$), -Z(Mn)  \\
	        & 685 & 681 & 683 & 686 &+Z(O$_1$), -Z(O$_2$)  \\
		  \hline\noalign{\smallskip}
	  E$_{2}$  & 85  &     & 83  &     &X,Y(Y$_1$, Y$_2$, Mn) \\
	           &     &     & 104 &     &+X,Y(Mn,O$_4$,O$_3$), -X,Y(Y$_1$, Y$_2$)  \\
	           & 142 & 135 & 137 &     &+X,Y(Y$_1$), -X,Y(Y$_2$) \\
%	           &     &     &     &     &+X,Y(Y$_2$), -X,Y(Y$_1$) \\
	           &     & 215 & 220 &     &+X,Y(O$_2$, Mn), -X,Y(O$_1$, O$_3$) \\
	           & 235 &     &     & 231 &Z(Mn,O$_2$,O$_1$) \\
	           & 249 &     &     &     &Z(Mn,O$_1$,O$_2$) \\	
	           & 309 & 302 & 305 &     &Z(O$_1$, O$_2$), +X,Y(O$_4$) \\
	           & 376 &     &     & 357 &+X,Y(O$_1$, O$_2$, O$_3$, O$_4$), -X,Y(Mn)  \\
	           & 418 &     & 405 &     &+X,Y(O$_1$, O$_4$), -X,Y(O$_2$, Mn) \\
	           & 442 &     &     & 441 &+X,Y(O$_4$), -X,Y(O$_1$ , Mn) \\
%		         &     &     &     & &+X,Y(O$_4$, O$_3$), +X,Y(O$_1$ , O$_2$) \\
%		         &     &     &     & &X,Y(O$_4$) \\
%		         &     &     &     & &X,Y(O$_4$, O$_3$) \\
		         & 637 &     &     & &X,Y(O$_3$, O$_4$) \\
	  \hline\noalign{\smallskip}
	  %\bottomrule
	 \end{tabular}
	\end{center}
	\label{freqphonon}
 \end{table*}

\begin{figure}
 \includegraphics*[width=9cm]{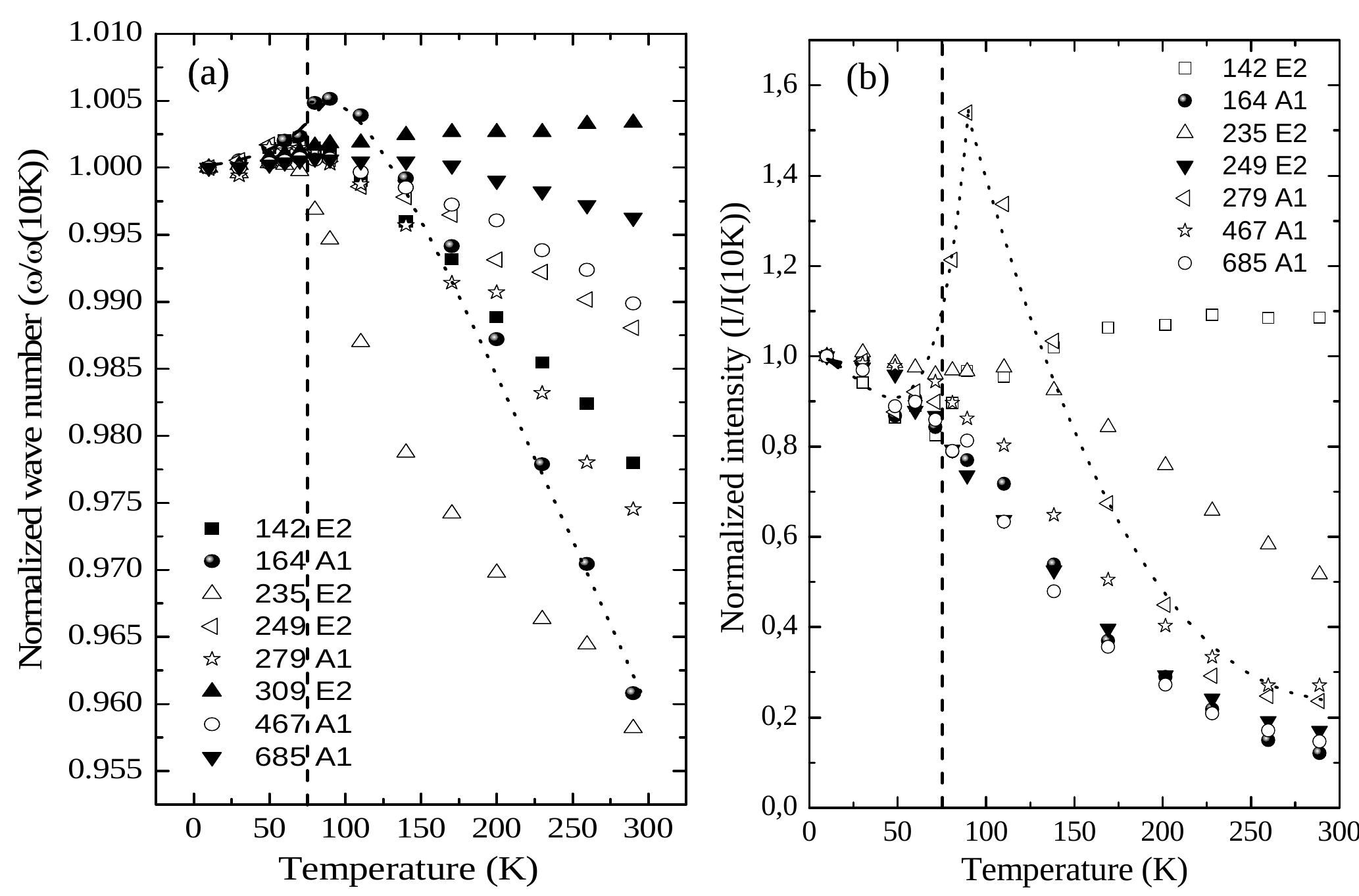}
 \caption{\label{Fig3} 	
a) Normalized wavenumbers ($\omega$(T)/$\omega$(10K)) and b) normalized intensities (I(T)/I(10K)) of  several \textsl{A}$_1$ and  \textsl{E}$_2$ modes as a function of temperature.}
\end{figure}

Figures \ref{Fig3}(a) and (b) show the normalized frequencies and the normalized intensities (over the values at 10 K) of several \textsl{A}$_1$ and \textsl{E}$_2$ modes.
The phonon frequencies usually tend to soften due to the dilation of the unit cell when temperature increases.
Except for the \textsl{E}$_2$ mode at 309 cm$^{-1}$, all frequencies are higher at low temperatures.
This mode is associated to the relative displacement of the apical oxygen ions along the c direction.\cite {Iliev1997}
It modulates the Mn-O-O-Mn bond and, hence, the super-superexchange Mn-Mn interaction between the adjacent Mn planes.
Remember that the dominant magnetic interaction is the Mn-O-Mn  antiferromagnetic superexchange within the planes whereas the Mn-O-O-Mn superexchange between neighboring planes is weaker by 2 orders of magnitude  {\cite{Lee, Petit}}. However, this latter interaction is involved in the stabilization of the three dimensional magnetic ordering below \textsl{T$_N$}.

The \textsl{A}$_1$ mode at 164 cm$^{-1}$ in Fig.~\ref{Fig3}(a) presents a frequency shift beyond the mean behavior of the other modes with a change of slope around the N\'eel temperature : a hardening at the magnetic transition followed by a softening at lower temperature.
This mode is related to the relative displacement of the apical yttrium ions along the c direction.
The measurements of the lattice parameters using high-resolution neutron diffraction have shown that the position of the \textsl{Y}$_2$ atoms along the z direction drops from 0.230 at 300 K to 0.2297 at 80 K just above T$_N$.\cite{Lee}
%Notice that, the role played by the R element in the RMnO$_3$ manganites has been underestimated. When R is a magnetic atom, several evidences show that R plays a direct role in the establishement of the magnetic state.\cite{Meier}However the role of non magnetic R atom is not negligeable as underlined by the temperature behaviour of the Y$^{3+}$ ions frequency at the the N\'eel temperature.

In Fig. \ref{Fig3}(b) the intensity of the \textsl{A}$_1$ mode at 279 cm$^{-1}$ strongly increases above T$_N$ and decreases below in contrast to the intensity of the other modes.
This mode is associated to the displacement of Mn ions in the a-b plane along the \textsl{x} axis.
It has been already shown that at T$_N$ \textsl{h}-YMnO$_3$ undergoes an isostructural transition with exceptionnaly large atomic displacement. In particular, the atomic displacement of the Mn ions is about 0.05-0.09$\dot{A}$ which is comparable to the values reported for prototype ferroelectric compounds.\cite{Lee2} The Mn ions shift away from the ideal of x=1/3 and the inplane Mn-Mn superexchange interaction is modified. Moreover the Mn-O bonds are no more equivalent (short one and long one) leading to the strong magneto-elastic coupling observed at T$_N$.
Therefore, the unusual frequency behavior of the \textsl{A}$_1$ modes around the N\'eel temperature is a fingerprint of the spin-phonon coupling in the magnetically-ordered phase.

\subsection{Magnetic excitations}

\begin{figure}
\resizebox{6.6cm}{!}{\includegraphics{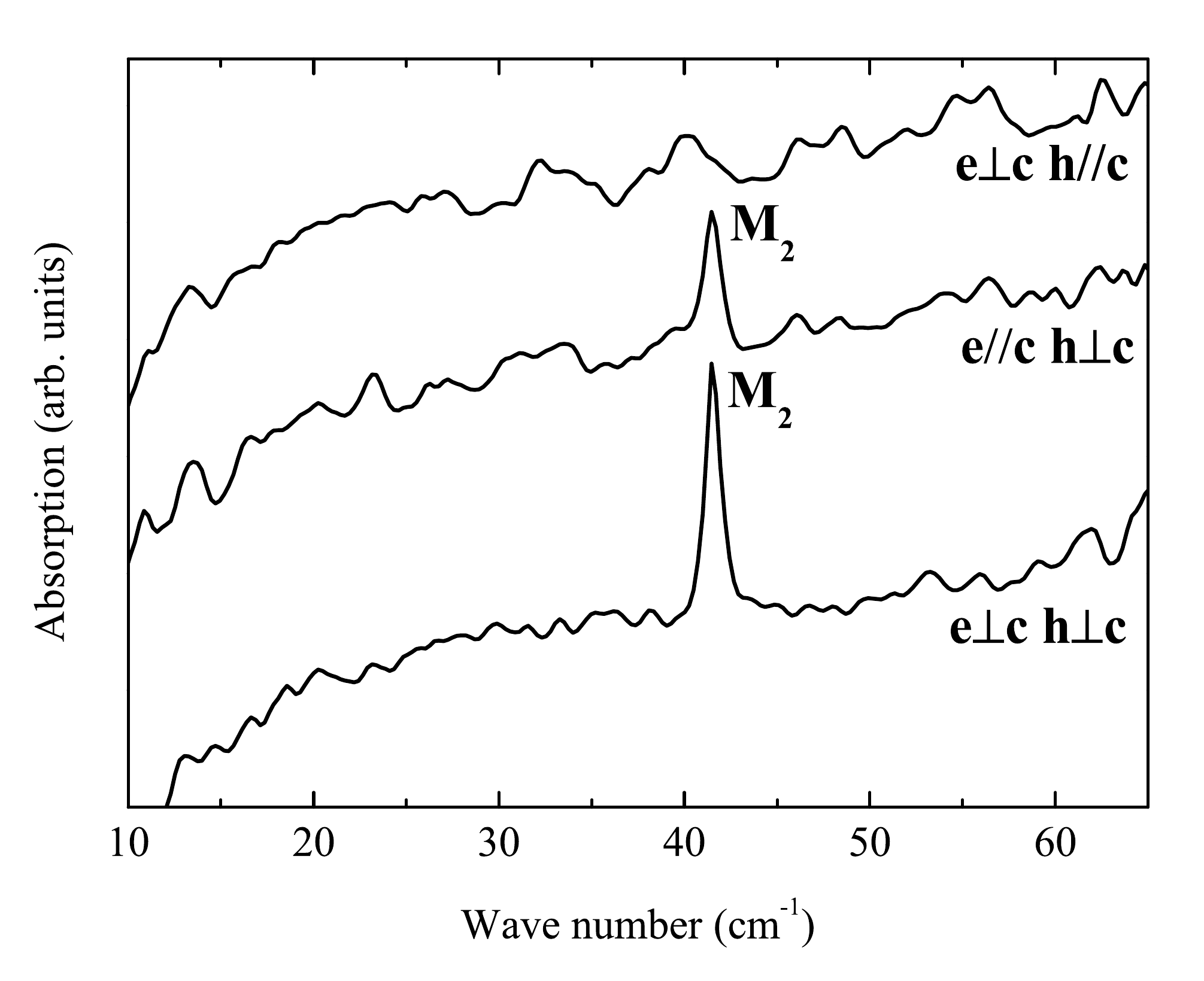}}
\caption{THz absorption spectra measured at 6K for the three different orientations of the THz electric and magnetic field as regards the $\bf{c}$-axis: {\bf e$\perp$c}~~ {\bf h//c},  {\bf e//c}~~{\bf h$\perp$c} and {\bf e$\perp$c}~~{\bf h$\perp$c}.}
\label{Fig4}
\end{figure}

\begin{figure}
\resizebox{6.6cm}{!}{\includegraphics{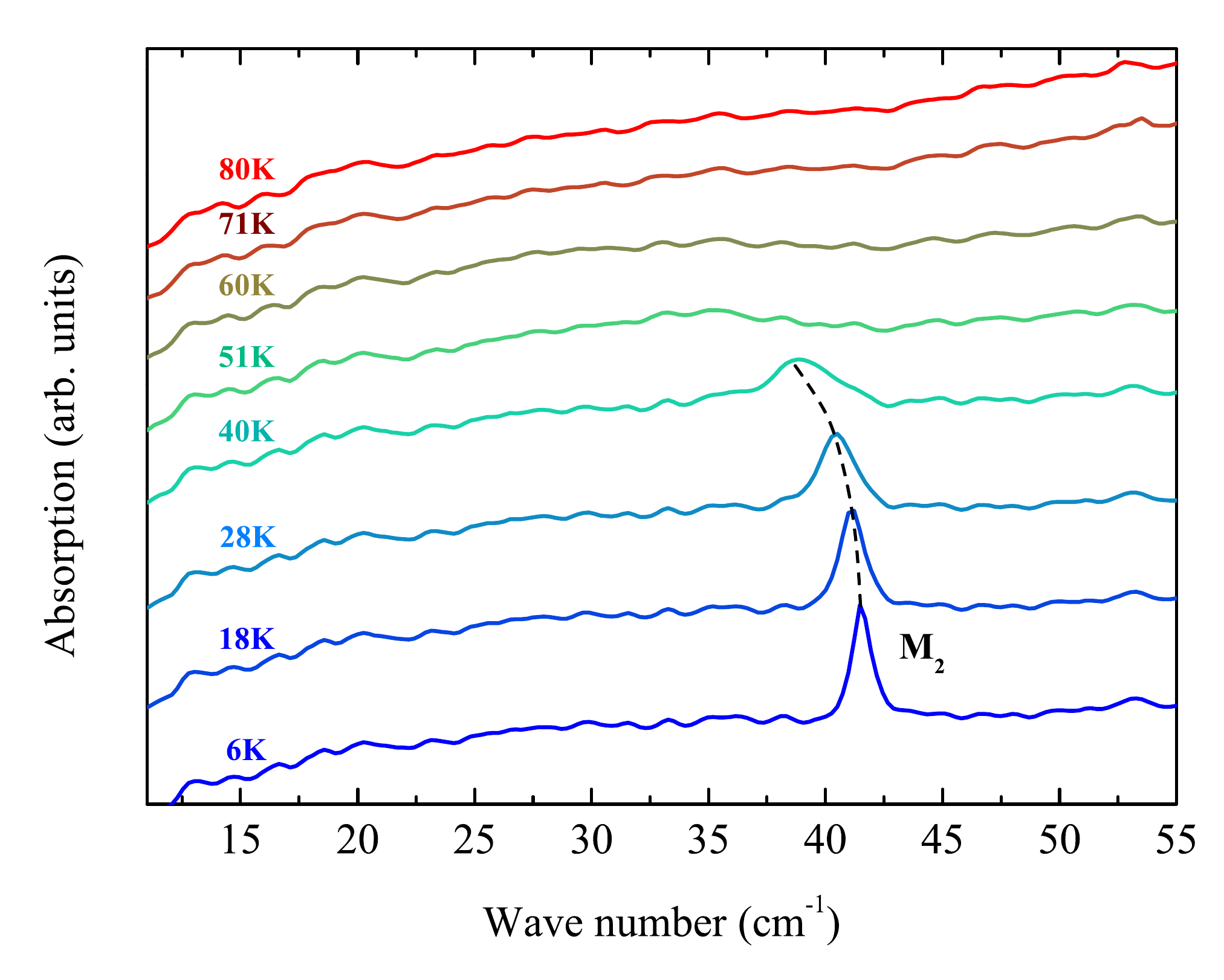}}
\caption{ THz spectra measured between 6 and 80 K for the {\bf e$\perp$c}~~{\bf h$\perp$c} selection rule.}
\label{Fig5}
\end{figure}

Magnetic excitations have been probed thanks to Raman as well as THz spectrocopies.
Figure \ref{Fig4} presents the measured THz spectra at 6K for all different orientations of the THz electric and magnetic fields with respect to the crystal $\bf{c}$-axis. One single excitation (labeled $\bf{M_{2}}$) is clearly observed at 41.5~cm$^{-1}$ whenever the THz magnetic field \textbf{h} is perpendicular to the $\bf{c}$-axis, that is to say for {\bf e//c}~~{\bf h$\perp$c} and {\bf e$\perp$c}~~{\bf h$\perp$c} while nothing occurs for {\bf e$\perp$c}~~{\bf h//c} (with \textbf{e} the electric field of the electromagnetic wave). It is therefore a magnetoactive excitation with {\bf h$\perp$c} as already reported in Ref.\cite{Kamba2011}. Its temperature dependence is given in Fig.\ref{Fig5}. Increasing the temperature has dramatic effect on $\bf{M_{2}}$: this excitation broadens and disappears above 60 K (below T$_N$), giving a clear indication that $\bf{M_{2}}$ is a magnon associated to the magnetic order.

\begin{figure}
\includegraphics*[width=7cm]{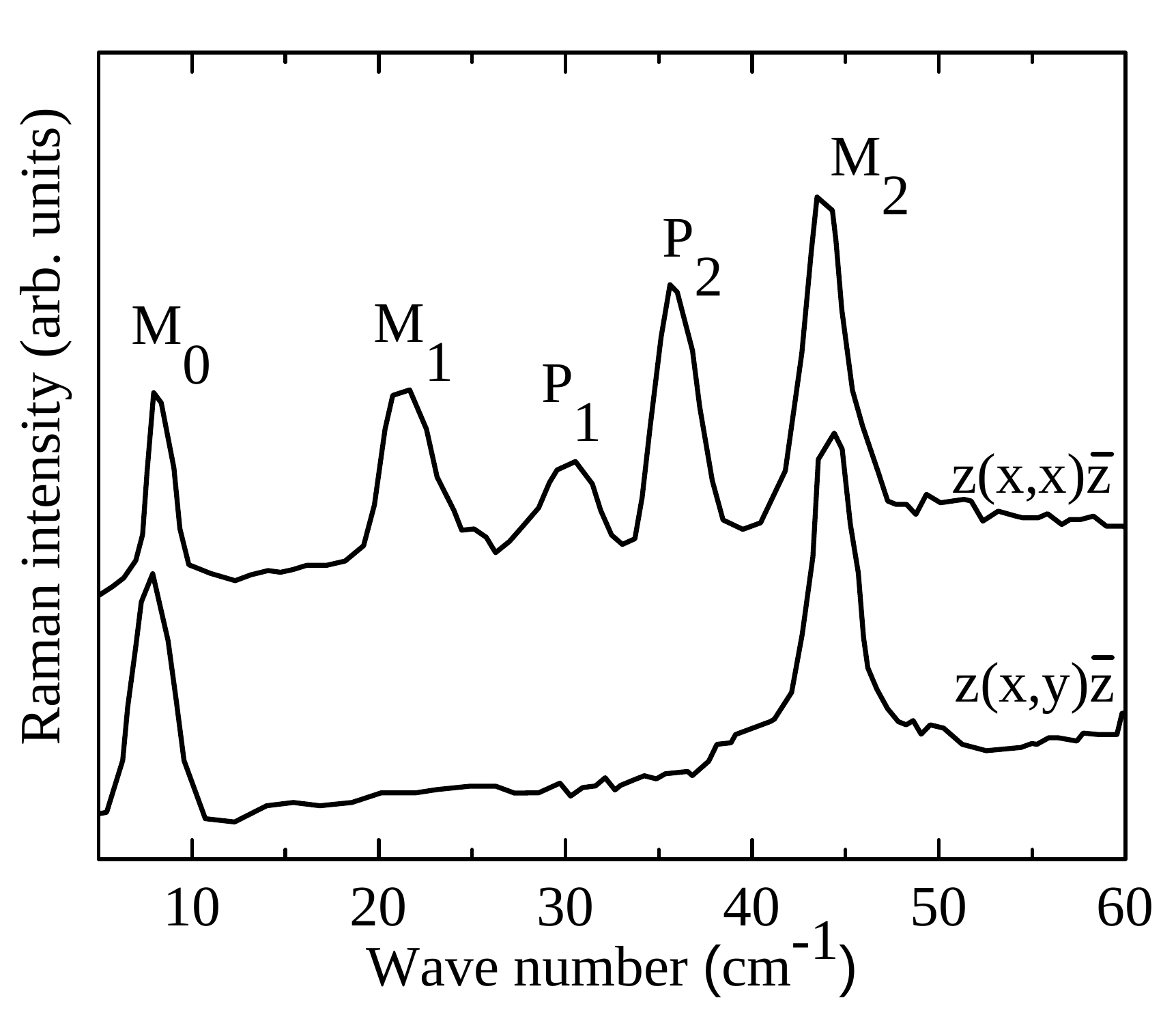}
\caption{	Raman spectra of low frequency excitations measured at 7 K in \textsl{z(xx)\={z}} and \textsl{z(xy)\={z}} configuration.}
\label{Fig6}
\end{figure}

Figure~\ref{Fig6} shows the low frequency Raman spectra measured on a single crystal at 7 K with two polarization configurations. Five peaks are detected: \textsl{M}$_0$ = 7.5 cm$^{-1}$, \textsl{M}$_1$ = 21 cm$^{-1}$, \textsl{M}$_2$ = 43 cm$^{-1}$, \textsl{P}$_1$ = 30.6  cm$^{-1}$ and \textsl{P}$_2$ = 35.4 cm$^{-1}$. All these peaks disappear above \textsl{T$_N$} (Fig.~\ref{Fig8}(a)) and are therefore connected to the magnetic order. Quite surprisingly, only one excitation, \textsl{M}$_2$, is observed both in Raman and THz experiments while the remaining ones are only Raman active. To understand the origin of these different observed excitations and their THz/ Raman activity, we now compare our results with those of neutron diffraction and spin wave calculations.

Inelastic neutron scattering measurements and their analysis have been reported previously \cite{Sato2003,Park2003,Petit}. Three different branches are observed for the magnons, that have been modelled using a Heisenberg spin hamiltonian with one or two antiferromagnetic interactions in the triangular planes and two interactions from one plane to the other. Planar and easy axis anisotropies have been included.
Note that both THz and Raman measurements can probe magnons at the zone center but with different selection rules and spectral weight. Magnetic excitations probed by neutron and THz measurements can be directly compared at the zone center since the interaction mechanism is the same: it is the magnetic interaction (between the magnetic moments in the sample with the one carried by neutron for the former, with the THz uniform magnetic field for the later). For Raman measurements, the interaction process is indirect (via spin-orbit coupling), so that the magnons energy  should coincide at the zone center or equivalent point in the reciprocal space, with no simple correspondence regarding their spectral weight. Note also that the optical measurements have a much better energy resolution than those with neutrons. Clearly, the numerous peaks, quite close in energy, observed in the optical spectroscopies, claim for more refined spin wave calculations. We have calculated the spin waves dispersion and spectral weight associated to the magnetic order within the $\Gamma_1$ irreducible representation in the linear approximation, taking into account the isostructural distortion occurring below $T_N$ \cite{Lee2}. We used the same spin Hamiltonian as described in Ref. \onlinecite{Petit}
 taking care that the Mn position in the triangular plane is shifted from the ideal 1/3 position to 0.03423 :

$\mathcal{H}=\sum_{R,i,R',j}J_{R,i,R',j}\overrightarrow{S}_{R,i}.\overrightarrow{S}_{R',j}+H\overrightarrow{S}_{R,i}.\overrightarrow{n}_{i}+DS_{R,i}^{z}.S_{R,i}^{z}$

where $\overrightarrow{S}_{R,i}$ denotes the spin at magnetic site i in the cell R, $\overrightarrow{n}_{i}$ is its mean direction unitary vector, $J_{R,i,R',j}$ describes the exchange interactions, while $H$ and $D$ correspond to easy-axis and easy-plane anisotropies, respectively. These parameters are refined using the inelastic neutron scattering measurements reported in Ref. \onlinecite{Sato2003,Park2003,Petit} and for more precision, our optical measurements (Raman and THz).

\begin{figure}
\resizebox{5.6cm}{!}{\includegraphics{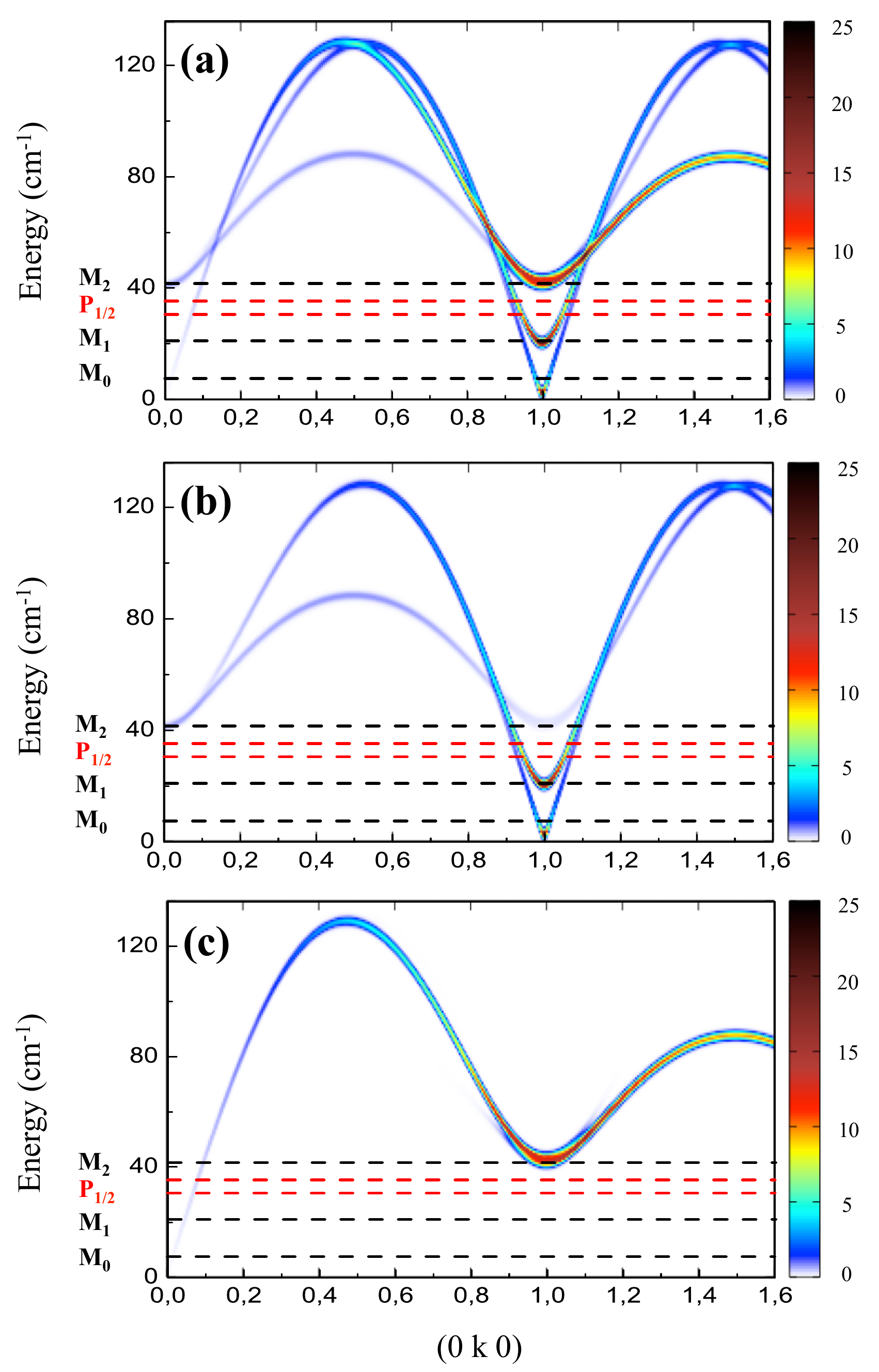}}
\caption{a) Spectral weight calculated for all spin components, b) for spin components perpendicular to the $\bf{c}$-axis, and c) along the $\bf{c}$-axis. Excitations, observed in Raman and THz measurements are reported on simulations.}
\label{Fig7}
\end{figure}

\begin{figure}
\includegraphics*[width=6cm]{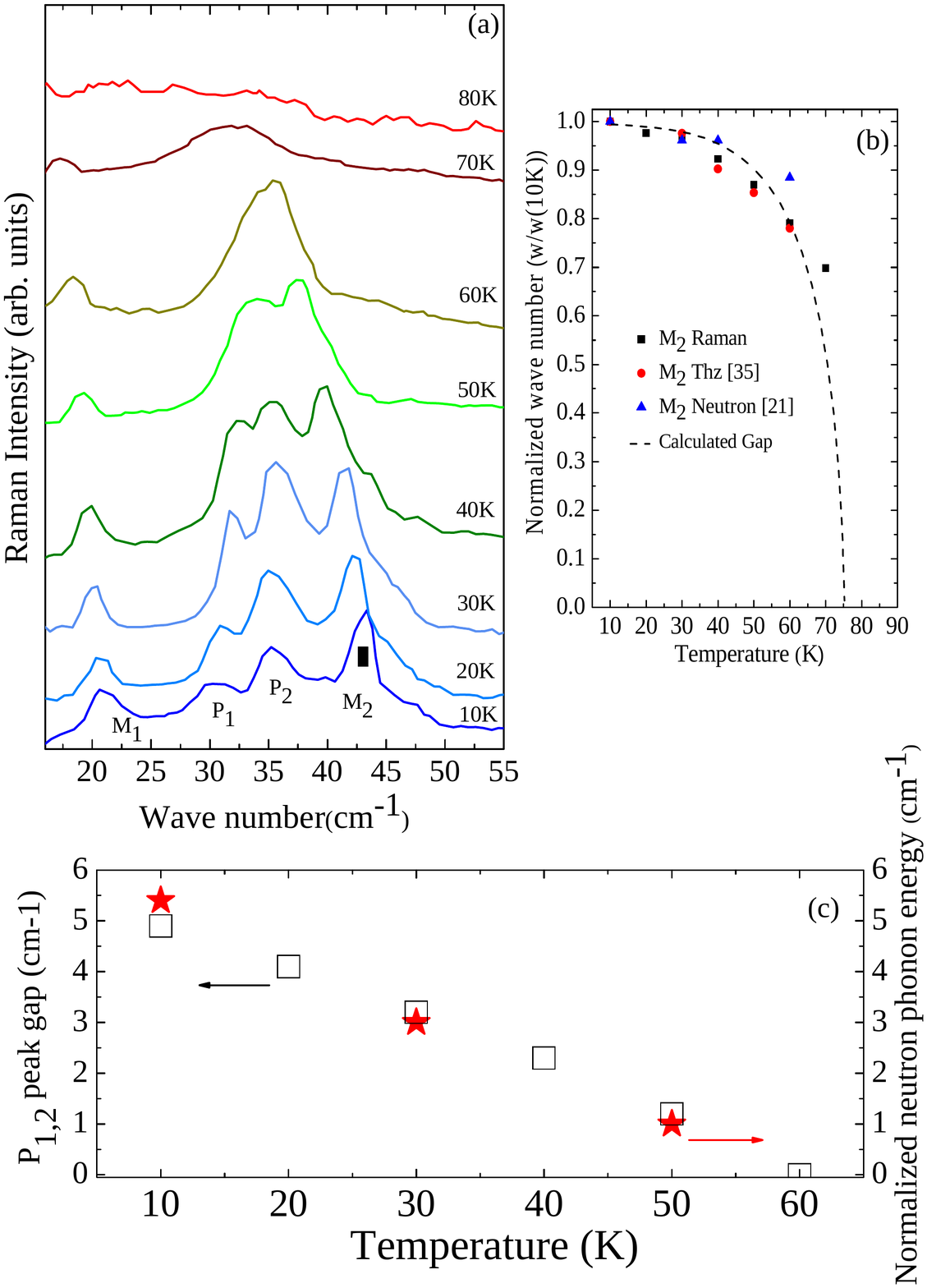}
\caption{a) Raman spectra measured between 10 and 80 K in \textsl{z(xx)\={z}} configuration; b) Temperature dependence of the \textsl{M}$_{2}$ peak energy measured by Raman (square), Thz (circle) in Ref. \onlinecite{Kamba2011} and neutron (triangle) in Ref. \onlinecite{Pailhes} compared to the calculated gap; c) Temperature dependence of the gap between the \textsl{P}$_{1/2}$ and the normalized phonon mode energy measured by Raman (square) and by neutron (star) at $q = 0.175$ in Ref. \onlinecite{Petit}. The normalization of the phonon mode energy corresponds to the difference between the phonon energy at a given temperature and its value at 200K.}
\label{Fig8}
\end{figure}

The best fit was achieved with the following exchange interactions (see Fig. \ref{Fig1}): $J_1=2.45$ meV is the average antiferromagnetic interaction in the triangular planes; $J_{z1}-J_{z2}=0.018$ meV is an effective interaction where $J_{z1}$ and $J_{z2}$ are the antiferromagnetic interplane interactions relative to the two different Mn-Mn interplane distances. As regards the anisotropies, we found $D=0.48$ meV for the easy-plane anisotropy that pushes the spins perpendicular to the c-axis and $H=0.0008$ meV the weaker easy-axis anisotropy within the easy plane. The results are plotted in Fig. \ref{Fig7}(a) for all the spin components, Fig. \ref{Fig7}(b) and (c) for spin components perpendicular and parallel to the c-axis. Three branches are generated with the following gaps at the zone center (0 0 0) or equivalent point (0 1 0): 2, 21 and 42~cm$^{-1}$.

Note the exchange of spectral weight when one moves along the (0 k 0) Brillouin zone direction. For instance, at the zone center (0 0 0), only one excitation with a finite energy around 42~cm$^{-1}$ has no vanishing spectral weight. This is the only excitation observed in THz measurements, ($M_2$), with the correct selection rule, {\bf h$\perp$c}, for spin components perpendicular to the c-axis.

Spectral weight and selection rules being different for Raman spectroscopy, there, all the three gaps are evidenced.
Notice that, the \textsl{z(xy)\={z}} configuration probes the magnon modes of the Mn$^{3+}$ magnetic structure perpendicular to the c-axis whereas the magnon mode in and out the (a,b) planes are measured using the \textsl{z(xx)\={z}} configuration. The energies of $M_1$ and $M_2$ excitations measured in Raman spectroscopy are in very good agreement with the spin wave calculations, whereas for $M_0$ some discrepancy persits. %The two magnon excitations M$_{2/3}$ illustrate the two different interplane interaction values J$_{z1}$ and J$_{z2}$ due to the weakly trimerization of the lattice the intratrimer and intertrimer Mn-Mn disctances of 3.42 and 3.62 $\dot{A}$, respectively.

To compare with more accuracy the Raman and neutron data, the normalized values of the \textsl{M}$_{2}$ peak as a function of the temperature are reported in Fig. \ref{Fig8}(b) in addition to the associated spin gap energy measured by neutron scattering\cite{Pailhes} and Thz spectroscopy\cite{Kamba2011} and the calculated energy gap. The temperature evolution of the spin excitation observed by Raman scattering is in good agreement with the behaviour of the magnetic moment of the Mn$^{3+}$ ions. The calculated gap energy is given by~: $E_{gap} = 2 S\sqrt{D J_{1}}$. $S$ follows the temperature behaviour of the magnetic moment measured in Ref. \onlinecite{Fabreges2010}.

%We also tested another spin hamiltonian with two different interactions in the triangular planes but we could not reproduce the $M_2$-$M_3$ splitting maintaining the constrain $J_{z1}>J_{z2}$ imposed by the $\Gamma_1$ irreducible representation \cite{Fabreges2010}.
Within these calculations, we explain the THz and Raman results for
 $M_1$, and $M_2$.
Clearly, there is no spin wave contribution that can explain $P_1$ and $P_2$. We now turn to possible lattice contribution. Notice that, no optical phonon mode is expected under 70 cm$^{-1}$.

Polarized inelastic neutron scattering measurements have reported the observation of a hybrid Goldstone mode \cite{Pailhes}. This hybridized mode results from the resonant interaction between an acoustic phonon branch and a magnon which is quite different from the hybrid modes called electromagnons resulting from the coupling between magnetic and optical phonon excitations.
The THz absorbtion measurements are able to reveal electromagnons at the zone center. Whereas conventional magnetic excitations are excited by the magnetic field component \textbf{h} of the THz wave,  electromagnons appear as magnetic resonances excited by the THz electric field  \textbf{e}. In Fig. \ref{Fig4}, there is no additional peak or modification as a function of the THz electric field and the only peak observed at 42 cm$^{-1}$ is driven by the THz magnetic field as expected for a pure magnetic excitation. From these measurements, it is clear that there is no evidence for electromagnons at the zone center in \textsl{h}-YMnO$_3$.

We now can come back to the  \textsl{P}$_{1,2}$  peaks in the Raman spectra of Fig.~\ref{Fig6}.
Based on our spin wave calculations, these excitations do not correspond to a one magnon process at the Brillouin zone center (Fig. \ref{Fig7}(a)).
To explain the origin of these peaks, several scenarii can be put forward:
%\par
%(1) The peaks might be associated with crystal field electronic transitions of the Y$^{3+}$ ions. However, since Y$^{3+}$ ($^1$S$_0$) has a complete 4$f$ shell and zero spin, its compounds should not have any Y-related crystal field electronic transitions.
\par
(1) the large phonon-paramagnon seen in Ref. \onlinecite{Kamba2011} corresponds in position. The Raman data resolve the \textsl{P}$_{1,2}$  peaks at this position, but they disappear at T$_N$ and we could not then associate them to this phonon-paramagnon excitation. The phonon-paramagnon is clearly visible at high temperatures in the FIR range that have not been explored in our THz study.
\par
(2) they could be associated to a two-magnons excitation with twice the energy  of the zone edge (around 16 cm$^{-1}$). %along the (0,0,k) direction However, the width of a two-magnons excitation is usally larger than the width of a one magnon excitation. Here the width of \textsl{P}${_4,5}$  is equal to  {the one of} \textsl{P}$_1$ (around 4 cm$^{-1}$).
However, there is no branch around 16 cm$^{-1}$ at the zone edges according to our spin wave calculations.
\par
(3) they might be associated to the anti-crossing between acoustic phonon and magnon dispersion curves.
Such anti-crossing has been already observed by neutron scattering around 40 cm$^{-1}$ and 60 cm$^{-1}$ for the lower and upper branches at the scattering vector $\textbf{q}_0 \approx 0.185\textbf{a}^{*}$ along the c-axis below the N$\acute{e}$el temperature. These values are not in total agreement with the energy of peaks \textsl{P}$_1$ at 30.6  cm$^{-1}$ and \textsl{P}$_2$ at 35.4 cm$^{-1}$. This discrepancy may come from the fact that the gap is not measured by neutrons and Raman at the same wave-vector and/or that the anti-crossing involves different acoustic phonon and magnon dispersion curves.

More features  support this kind of interpretation.
The temperature dependence of  peaks \textsl{P}$_{1,2}$ is reported in Fig. \ref{Fig8}. It is clear that they appear below T$_N$ and are then connected to the magnetic order.
In Fig. \ref{Fig8} (c), the energy difference between them is compared to the anti-crossing gap value measured by neutron measurements\cite{Petit} as a function of temperature. Both their values and temperature dependencies are similar.
Another argument is given by the Raman selection rules. The \textsl{P}$_{1/2}$  peaks appears only in the \textsl{z(xx)\={z}} configuration i.e. along the c-axis, the direction of the observed gap in neutron scattering.
These evidences support the interpretation of \textsl{P}$_{1/2}$  peaks as  signatures in  Raman spectra of the anti crossing between a magnon and an acoustic phonon branches.

One question remains: how is it possible to observe such a gap by the way of a Raman scattering process?
Raman scattering probes all the dispersion curve of an excitation through a two scattering process involving twice this excitation with +q and -q wavevectors.
Such a process gives rise to a Raman signal with an intensity proportional to the density of state. A flat region in the dispersion curve and the associated strong density of state give a Raman peak that can be measured. This is the case for the two magnon modes of the Brillouin zone edge.
A gap in a phonon dispersion corresponds also to this criteria. However, in a two scattering process, the Raman peaks are observed at twice the energy of the excitation. %In Fig. \ref{Fig6}, \textsl{P}$_{1,2}$  peaks are observed at energies lower than those observed with neutrons at $\textbf{q}_0$ where the acoustic gap opens. %To explain the Raman activity, the Brillouin zone has to be folded at a lower  $q$.

\begin{figure}
 \includegraphics*[width=6cm]{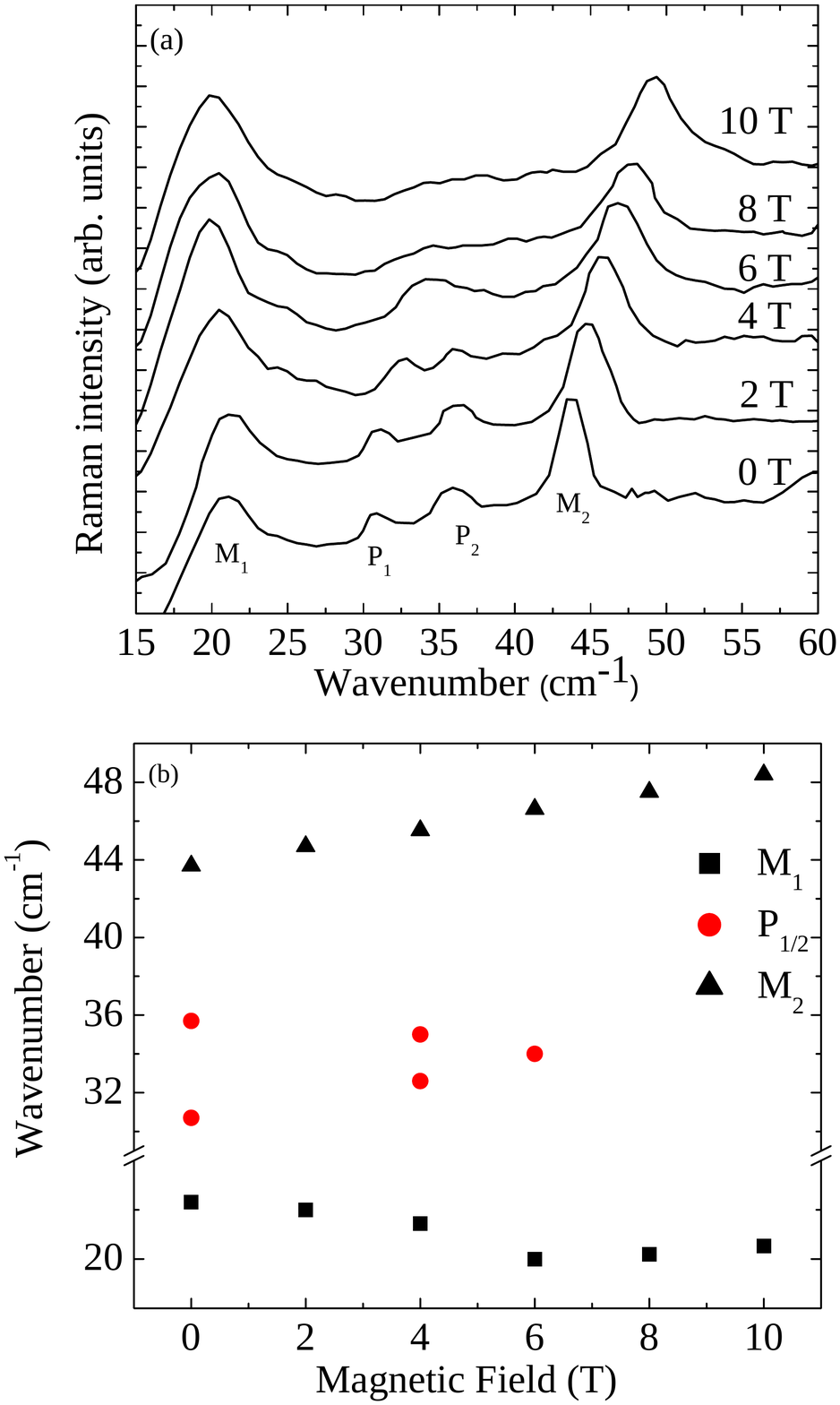}
 \caption{\label{Fig9} 	
a) Raman spectra obtained at 10 K in \textsl{z(xx)$\bar{z}$} configuration using a magnetic field along the \textsl{c} axis, b) wavenumber of \textsl{M}$_1$, \textsl{M}$_2$, and \textsl{P}$_1/2$ excitations as a function of the magnetic field.}
\end{figure}

To shed some light on the impact of the magnetic structure on the spin excitations, we have investigated the magnetic phase diagram of \textsl{h}-YMnO$_3$.
The frequencies of the magnetic excitations are reported as a function of the applied magnetic field along the c axis in Fig. \ref{Fig9} (b).
The frequencies of the \textsl{M}$_1$ excitation is almost constant, whereas  \textsl{M}$_2$ peak  increases in frequency.
 %These modes are associated with theeasyplane anisotropies.%
No phase transition is detectable when applying a magnetic field up to 10 T. This is in contrast with other hexagonal manganites \textsl{h}-RMnO$_3$ (R=Ho, Er, Tm, Yb) where a reordering of the magnetic structure have been observed for different values of the applied magnetic field due to the coupling between the Mn$^{3+}$ ions and the magnetic rare earth.\cite{Fiebig2003} \textsl{h}-YMnO$_3$ (\textsl{h}-ScMnO$_3$ or \textsl{h}-LuMnO$_3$) has a complete 4f shell and therefore doesn't display an antiphase rotation of the Mn$^{3+}$ spins when a magnetic field is applied. Thus in \textsl{h}-YMnO$_3$, no magnetic ordering transition is expected with the magnetic field.
%In an external magnetic field $H$ along the c-axis, the doubly-degenerate magnon (already splitted thanks to the easy plane anisotropy) splits further into two branches according to its effective g-factor and follows the equation: $\omega^{\pm}_M(B)=\omega^{\pm}_M(0T)\pm1/2g\mu_BH$ where $\mu_B$ is the Bohr magneton. The fit of the field dependant magnon  frequencies gives a g-factor equal to $2.10\pm0.1$. This value is in agreement with the one expected for isolated Mn$^{3+}$ ions with a weak spin-orbit contribution.
The $P_{1/2}$ excitations  are sensitive to the magnetic field: they merge and widen to disappear above 6 T. This confirm that they have  a magnetic origin, at least partially. Further characterization is required to understand fully these new excitations.

\section{conclusion}

In summary, the detailed study of the lattice excitations in \textsl{h}-YMnO$_3$ confirms the strong magneto-elastic coupling in this compound.
The pure magnetic excitations observed by Raman and THz spectroscopies are consistant with the neutron calculations and measurements. No evidence for electromagnons has been found in  \textsl{h}-YMnO$_3$. Two unexpected low frequency excitations have been measured. We compare these excitations with  the anti-crossing between a magnon and an acoustic phonon branches. The Raman activation of such excitations needs to be understood but it underlines the strong spin-phonon coupling in this compound.

%that the role of the super-superexchange Mn-Mn interaction can not be rule out in the magnetic phase transition.Moreover our measurments underline the role played by the R element in the RYMnO$_3$ manganites.
% Raman and THz experiments provide evidence for a strong coupling between spins and phonons with the opening of a gap below T$_N$ in the dispersion of an acoustic mode.
% However, this coupling does not lead to the creation of an hybridized electromagnon excitation.

\section*{Acknowledgments}
This work was supported in part by the French National Research Agency (ANR) through DYMMOS project, the General Directorate for Armament (DGA). We thank J. Debray for the samples preparation and orientation.

\end{document}